\documentclass[%
nofootinbib,
amsmath,amssymb,
aps,
prd,
longbibliography,
]{revtex4-2}

\usepackage{xcolor}
\usepackage{graphicx}
\usepackage{dcolumn}
\usepackage{bm}
\usepackage[mathlines]{lineno}

\usepackage[hyperfootnotes = false]{hyperref}
\usepackage{cleveref}
\hypersetup{
 colorlinks,
 linkcolor=blue,
 citecolor=red,
 urlcolor=blue, 
 }

\begin{document}

\newcommand{\todo}[1]{{\textcolor{red}{{#1}}}}

\newcommand{\change}[1]{#1}

\def\be{\begin{equation}}
\def\ee{\end{equation}}
\def\beq{\begin{equation}}
\def\eeq{\end{equation}}
\def\bea{\begin{eqnarray}}
\def\eea{\end{eqnarray}}

\def\bear{\begin{eqnarray}}
\def\eear{\end{eqnarray}}

\def\lna{{\hat a}}
\def\lnb{{\hat b}}
\def\lnc{{\hat c}}
\def\lnd{{\hat d}}
\def\lne{{\hat e}}
\def\lni{{\hat i}}
\def\lnj{{\hat j}}
\def\lnk{{\hat k}}
\def\lnm{{\hat m}}

\def\lnZ{{\hat 0}}
\def\lnO{{\hat 1}}
\def\lnD{{\hat 2}}
\def\lnT{{\hat 3}}

\def\ta{{\tilde{a}}}
\def\tj{{\tilde{j}}}
\def\tn{{\tilde{n}}}
\def\tp{{\tilde{p}}}
\def\ts{{\tilde{s}}}
\def\tu{{\tilde{u}}}
\def\tx{{\tilde{x}}}

\def\tB{{\tilde{B}}}
\def\tE{{\tilde{E}}}
\def\tF{{\tilde{F}}}
\def\tJ{{\tilde{J}}}
\def\tT{{\tilde{T}}}
\def\tY{{\tilde{Y}}}

\def\F{{\mathcal F}}

\def\tbeta{{\tilde{\beta}}}
\def\teps{{\tilde{\varepsilon}}}
\def\tsig{{\tilde{\sigma}}}
\def\ttheta{{\tilde{\theta}}}
\def\tvort{{\tilde{\omega}}}
\def\tmu{{\tilde{\mu}}}
\def\tnu{{\tilde{\nu}}}
\def\tom{{\tilde{\omega}}}

\def\tperp{{\tilde{\perp}}}

\def\veps{{\varepsilon}}
\def\eps{{\epsilon}}

\newcommand{\parallelsum}{\mathrel{/\mkern-5mu/}}

\newcommand{\la}{{\langle}}
\newcommand{\ra}{{\rangle}}

\title{Higher-level large-eddy filtering strategy for general relativistic fluid simulations}

\author{T. Celora$^1$, N. Andersson$^2$, I. Hawke$^2$, G.L. Comer$^3$ \& M.J. Hatton$^2$}
 
\affiliation{%
$^1$ Institute of Space Sciences (ICE, CSIC), Campus UAB, Carrer de Magrans, 08193 Barcelona, Spain\\
$^2$ Mathematical Sciences and STAG Research Centre, University of Southampton,
Southampton SO17 1BJ, United Kingdom\\
$^3$ Department of Physics, Saint Louis University, St.~Louis, MO,
63156-0907, USA}

\date{\today}

\begin{abstract}
Nonlinear simulations of neutron star mergers are complicated by the need to represent turbulent dynamics. As we cannot (yet) perform simulations that resolve accurately both the gravitational-wave scale and the smallest scales at which magneto/hydrodynamic turbulence plays a role, we need to rely on approximations. 
Addressing this problem in the context of large-eddy models, we outline a coherent Lagrangian filtering framework that allows us to explore the many issues that arise,  linking conceptual problems to practical implementations and the interpretation of the results. 
We develop understanding crucial for quantifying  unavoidable uncertainties in current and future numerical relativity simulations and consider the implications for neutron-star parameter estimation and constraints on the equation of state of matter under extreme conditions. 
\end{abstract}

\maketitle


\section{Motivation\label{sec:level1}}

Compact binary mergers involving neutron stars---be it with another neutron star or a black hole---are generally considered exciting laboratories for physics under extreme conditions \cite{baiotti_binary_2017}. This notion was made sharp reality with the first binary neutron star merger (GW170817) observed by the LIGO-Virgo instruments, the subsequent gamma-ray flash caught by the Fermi and Swift satellites and the emission across the observable electromagnetic band that followed~\cite{abbott2017gw170817,LIGOScientific:2017ync}. The amount of information gleaned from this single event is truly amazing and the promises of future multimessenger astronomy are inspirational.
In particular, gravitational-wave observations and multimessenger detections have the potential to constrain microphysical parameters of matter at supranuclear densities and under extreme pressures. However, to get the most out of such observations, signals need to be compared to precise quantitative models. For scenarios such as binary neutron-star mergers (at least part of) the model prediction requires nonlinear numerical simulations. This motivates the ongoing effort to improve the simulation technology, adding  as much realism as possible to the models \cite{dionysopoulou2013general,wright_resistive_2018,Cipolletta2019Spritz,wright_resistive_2020,hammond2021thermal,chabanov2021viscous,radice2022newM1,sun2022jetM1+B,werneck2023illinoisGRMHDextension,foucart2023neutrino,espino2023GRMHDuncertainties,BVinSim,zappa2023BNSnuComparison,cook2023athena++,kiuchi2023SecondLongBNS,Hatton2024DEIFY,espino2024neutrino,most2024emergence,Kalinani2024asterX}.

A key aspect of the merger problem is  turbulence.  The merger process is sufficiently nonlinear to drive the neutron star fluid into the turbulent regime. A Kelvin-Helmholtz-like instability is triggered as the two stars touch \cite{priceRosswog2006,Kiuchi2015_KHI} and the star's magnetic field coupled to the fluid rotation may drive the magneto-rotational instability \cite{palenzuela2022turbulent,celora2024MRI,siegel2013magnetorotational}, as well. In effect,  one would expect significant power and information to be present on all length-scales, from the size of the system down to the dissipation scale. This is ``problematic'' given that the best current estimates of the dissipation scale (centimeters or smaller) are many orders of magnitude below what is practical to numerically simulate (tens of meters).

As any information below the numerical grid scale is lost in a simulation, the best one can do is to replace the ``true'' model (which is correct all the way down to the dissipation scale) with an ``effective'' turbulence model. 
This strategy typically involves a filtered/averaged version of the dynamical equations.
As filtering leads to residual terms when applied on any nonlinear expression, the filtered equations are completed by some phenomenological ``closure'' scheme intended to represent the ``missing'' small-scale dynamics \cite{DavidIanReview}.
Ultimately, the development of such Large Eddy Schemes (LES) aim to provide  models that lead to ``more physical'' results when used at practical, numerically accessible, scales. 

Current efforts in this direction recognise the ``implicit'' filtering associated with the numerical discretisation and the ``explicit'' filtering associated with a chosen closure model. Both aspects impact on the simulation results, but it may not be practical---at least not at this point---to try to figure out all implicit issues. Some features of the problem are simply unavoidable.  Implicit filtering is i) associated with the grid size, and ii) will damp high-frequency behaviour for stability. We also know that,  iii) stable numerical schemes involve some kind of entropy argument. We will comment on these issues later, at which point it will become evident that their impact is not easy to disentangle. We can, however, make an effort to understand/improve the  tools required to deal with explicit filtering. This is the motivation for the present discussion.

\section{Setting the stage}

For later convenience, let us begin by stating the problem we would like to solve. Neutron star mergers involve the nonlinear dynamics of extreme density matter moving at high velocity and reaching impressive temperatures, in a strong gravity setting. In addition, we likely have to consider the impact of a strong magnetic field and aspects associated with neutrinos. There is much about the problem that we do not yet understand---at least at the level  of detail required to match results against future, precision observations---but it is relatively straightforward to write down the equations we need to consider. 

Assuming that we restrict ourselves to Einstein's theory of gravity, the starting point will be the standard field equations (in geometric units, $c=G=1$ and using $a,b,c,\ldots$ to represent spacetime indices)
\begin{equation}
    G_{ab} = 8\pi T_{ab} \;,
\end{equation}
linking the spacetime curvature (expressed in terms of the Einstein tensor $G_{ab}$) to the stress-energy(-momentum) tensor ($T_{ab}$) of matter. A realistic matter description would have to account for a number of aspects, starting from the matter composition,  thermal effects, neutrinos,  the state of matter (a mature neutron star will have an elastic crust and probably superfluid/superconducting components in its interior) and so on. This is a  complex problem \cite{LivRev_NilsGreg}. 

A thorough discussion would take us beyond what is required for the points we want to stress, so we ignore many of the relevant physics aspects and focus our attention on the standard perfect fluid model for which we have
\begin{equation}
    T^{ab}_{\text{F}} = (\veps+p) u^a u^b + p g^{ab} = \veps u^a u^b + p \perp^{ab} \;.
    \label{Tpf}
\end{equation}
Here $u^a$ is the fluid four-velocity, $p$ is the (isotropic) pressure and $\veps$ is the energy density. The spacetime metric is $g_{ab}$ and 
\begin{equation}
\perp^{ab} = g^{ab} + u^a u^b \;,
\end{equation}
acts as a projection orthogonal (in the spacetime sense) to $u^a$.

The equations need to be supplemented by an equation of state, ideally provided by (or at least compatible with) nuclear physics arguments. The simplest model would be a cold, barotropic (single-fluid) model expressed in terms of the baryon number density, $n$. The fluid pressure, $p$, then follows from 
the thermodynamic  relation (effectively, the first law)
\begin{equation}
    p + \veps = n\mu \;,
    \label{Gibbs}
\end{equation}
where the chemical potential is given by
\begin{equation}
    \mu =\frac{d\veps}{dn}\;.
\end{equation}
In essence, all relevant fluid quantities follow from $\varepsilon(n)$.
These relations are easily extended to finite temperature, multi-component systems \cite{LivRev_NilsGreg}, and corresponding tabulated equations of state can be found, for example, in the compOSE database \cite{compOSE}. However, the single-parameter model will suffice for our present purposes.

In terms of dynamical equations, the flow must satisfy baryon number conservation so we need to ensure that
\begin{equation}
    \nabla_ a n^a = 0 \;,
    \label{baryon}
\end{equation}
where the baryon number flux is $n^a = n u^a$.

Adding electromagnetism to the problem, 
it is natural to work with the Faraday tensor, $F^{ab}$, in terms of which the Maxwell equations take the form
\begin{equation}\label{eq:MaxwellCovariant}
    \nabla_aF^{ba} = \mu_0 j^b \;, \qquad \nabla_{[a}F_{bc]} = 0 \;,
\end{equation}
with $j^a$ the charge (four-)\,current.
Assuming minimal coupling to the matter, we have
\begin{equation}
    T^{ab}_\text{EM} = \frac{1}{\mu_0}  \left[ F^{ad}F^b_{\phantom{b}d} - \frac{1}{4} g^{ab}F^{cd}F_{cd}\right]\;,
\end{equation}
with $\mu_0$ the magnetic permeability.
The fluid equations of motion are then obtained from (given that $\nabla_a G^{ab}=0$ by the Bianchi identities)
\begin{equation}\label{eq:MHDproxyFluid}
	\nabla_a T^{ab}_{\text{F}} = - \nabla_aT^{ab}_\text{EM} = -j_a F^{ab}\;,
\end{equation}
where the term on the right-hand side represents the Lorentz force. 
The system of equations is closed provided the charge current is linked to the remaining variables. This link is ideally provided by an underlying multi-fluid model \cite{LivRev_NilsGreg}---appropriately representing the physics involved---although for practical applications it is commonly expressed in terms of a phenomenological Ohm's law \cite{andersson2022physics}.

In astrophysical applications, it is customary to work with the---observer dependent---electric and magnetic fields. Taking the observer to be associated with the fluid flow (for now), we have
\begin{equation}
    F_{ab} = 2u_{[a}E_{b]} + \veps_{abc}B^c  \;,
    \label{eq:FaradayDecomp}
\end{equation}
with $\veps_{abc} = \veps_{dabc}u^d $. This then leads to 
the electric field
\begin{equation}
    E_a =  F_{ab}u^b \;,
    \label{efield}
\end{equation}
and the magnetic field
\begin{equation}
    B_a =  \frac{1}{2} \veps_{abc}F^{bc} \;.
    \label{bfield}
\end{equation}
Finally, we may decompose the charge current as
\begin{equation}
    j^a = \sigma u^a + J^a \;,
\end{equation}
with $u_aJ^a=0$ and $\sigma$ the charge density measured by the chosen observer. 

In addition, many astrophysical systems are expected to be well represented by ideal magnetohydrodynamics (MHD). Because of the high conductivity, the local electric field can be taken to vanish ($E_a=0$) and the matter is charge neutral ($\sigma = 0$). Charge neutrality is, in fact, expected to hold on the nuclear physics scale so it would seem natural---given that the properties of a given fluid element are assumed to be homogeneous and isotropic---to retain this assumption in a frame comoving with the fluid.
Under this assumption, the electric field loses its role as independent variable, instead it must be determined by some form of Ohm's law, thus considered as a closure relation for the electric field rather than for the charge current \cite{andersson2022physics}.

So far, so good. We have stated the equations we need. All we have to do is solve them and the job is done. The problem is that we cannot easily do this. We need to resort to numerical simulations which, in turn, necessarily rely on approximations. This is where the trouble starts.

When constructing a simulation of relativistic matter, the starting point will always be the equations of motion.  For most models one directly imposes stress-energy conservation, and as many additional dynamical equations (for example, conservation of particle number) as needed. When simulating problems with shocks (such as neutron star mergers) it is important to use a balance-law form, which is straightforward for the equations given by stress-energy conservation. This, in turn, introduces a set of variables---the \emph{conserved} variables---that are evolved. To complete the system we need constitutive relations (such as the equation of state) and  (algebraic) relations that link the conserved variables to any other variable we may need (for example, to compute the temperature needed as input for the equation of state). Intuitively, it is at the level of these \emph{primitive} variables that a connection is made to the nuclear physics.

The steps just described are generic enough to hold for most problems of interest. However, current simulations typically rely on additional simplifications, effectively breaking the mathematical link between the physical quantities. This is particularly noticeable when comparing the conservative to primitive conversion for the ideal fluid case \cite{Kastaun_iMHD_c2p,Siegel_MHD_c2p,Galeazzi_iHD_c2p}, where the procedure requires solving a single nonlinear scalar equation, to more general cases such as the Israel-Stewart approach to dissipative fluids \cite{DelZanna_2013,Du_Heinz_2020,shibata2017viscous} or even elastic matter \cite{GHE_2012}, where the procedure requires solving  systems of nonlinear equations. The key point is---as we shall soon see---that the relations we need to invoke are  nonlinear. 

\section{Quantifying uncertainties}

The discussion in this paper can be considered at different levels and from several perspectives. Broadly speaking, we are discussing a Lagrangian based approach to the filtering of turbulent relativistic fluid dynamics---see \cite{Celora2024lagrangian} for a recent practical demonstration of the strategy. From a conceptual perspective, our discussion also raises  questions concerning \emph{uncertainty quantification}. Here we have in mind the following steps (formalised in, for example, \cite{UQchapter}), relevant for observational parameter extraction. In the modelling process we specify an initial-boundary value problem where some aspects (such as the microphysical closure relations, including the equation of state) are to some extent uncertain. The \emph{forward} uncertainty quantification propagation process computes the impact of the uncertainties in the model specification onto uncertainties in the observable signals, through a combination of numerical simulations and approximate models. The statistical distributions that result are then combined with observational uncertainties, for example due to instrument noise, to perform \emph{backward} uncertainty quantification when using the data to perform parameter extraction.

These steps are standard. However, there are more uncertainties relevant for uncertainty quantification than  the ``true'' microphysical closures. Numerical simulations already consider the (hopefully subdominant) uncertainties due to discretisation error and issues with initial-boundary conditions. What is \emph{not} usually considered are the uncertainties in the equations of motion induced by the conceptual issues from filtering. If the same filtering residual can be interpreted either as an effective viscosity term or as a correction to the microphysical closures---as we, indeed, argue in the following---then it introduces systematic uncertainties that need to be considered in the parameter estimation process.

One of the core purposes of this work is to show how different choices in the construction of the filtered model lead to differing interpretations of ``the same'' closure relations and their associated ``microphysical'' parameters. As such choices are unavoidable, their consequences must be considered in assessing the accuracy of signal predictions from numerical simulations, so that the resulting uncertainties can be folded into the parameter extraction effort. Detailed calculations to do this are not possible until a more precise framework is laid out. This paper is a first step in developing such a framework.

\section{Establishing the language}

In order to avoid confusion, it is helpful to establish the terminology,
notation and definitions we need in order to make the discussion precise.
First of all, we need to introduce a \emph{micro-model}. This is a matter model defined at all scales. It is computable, with no free coefficients. In essence, everything is known. Our default micro-model---mainly for demonstrative purposes---will be the relativistic ideal fluid defined via the stress-energy tensor \eqref{Tpf} alongside the continuity equation \eqref{baryon}.
The micro-model also specifies the pressure using the equation of state (via thermodynamical arguments, as before) which is assumed to be completely known. From a mathematical perspective, the equation of state represents a \emph{closure relation} that ensures that we have all the information we need to solve the problem. It could be provided by detailed nuclear physics arguments or a pragmatic model like the simple gamma-law $p = (\gamma - 1) n e= (\gamma-1)(n-\veps)$, where $e$ is the specific internal energy and $\gamma$ is assumed to be known. The  details do not matter. The arguments we will make hold in general.

Whilst we assume the micro-model to be completely known, we also need to admit that this is not the model that will actually be used (for example in a numerical simulation). This is inevitable given that the fluid behaviour at small scales cannot be practically captured. Instead, we have to use a \emph{meso-model}. In essence, the meso-model is a \emph{class} of models where some of the coefficients, or (again) closure relations/residuals, are assumed initially unknown and will be \emph{chosen} in order to best match (in some suitable sense) the micro-model. Our default meso-model will be the relativistic \emph{non-ideal} fluid defined via the stress-energy tensor (the tildes will be explained later)
\begin{equation}
    \tilde{T}_{ab} = \tilde{\varepsilon} \tilde{u}_a \tilde{u}_b + \tilde{p} \tilde{\perp}_{ab} + \tilde{\tau}_{ab} \;,
    \label{Tni}
\end{equation}
along with the continuity equation
\begin{equation}
    \tilde{\nabla}_a \tilde{n}^a = \tilde{\Gamma} \;,
    \label{tdna}
\end{equation}
for the particle number flux. 

What difference does it make?
There are (at least) four points where the details of the meso-model can be modified to match the micro-model. One is in the choice of the equation of state, which (for example) links the effective pressure, $\tilde{p}$, to the other thermodynamic potentials in the meso-model ($\tilde n$ and $\tilde \varepsilon$, say). It seems intuitive that this will be tightly linked to the equation of state from the micro-model, but there is no in-principle reason why they should be identical. Second, we have to fix the non-conservative $\tilde \Gamma$ term  in the continuity equation, an issue related to the observers chosen in the micro- and meso-models. Again, it seems intuitive that $\tilde \Gamma$ ought to vanish, but there is no in-principle guarantee of this. Third, we have to deal with the non-ideal stresses, $\tilde{\tau}_{ab}$. These can be used to capture effects that appear to be anisotropic once the smallest scales (captured by the micro-model) are neglected (in the meso-model). Finally, there is the spacetime metric.

The notation we use in the following is such that all variables in the meso-model that are conceptually related to variables in the micro-model have a tilde; so $\tilde{\varepsilon}$ ($\varepsilon$) is the energy in the meso- (micro-) model as seen by the  fluid observer $\tilde{u}^a$ ($u^a$). These quantities will be related by a filtering operation, schematically represented by $\langle \cdot \rangle$, but it is notably \emph{not true} that $\tilde{\cdot} = \langle \cdot \rangle$, in general. As we will demonstrate in the following, the \emph{choice} of precisely which variables are directly related via the filtering operation is crucial in determining its impact.

\section{State of the art}

Let us now turn to the matter at hand: large-eddy models for general relativistic fluid simulations.
Recent interest in this problem has been driven by the understanding that  essential physics occurs on length scales that cannot be captured in nonlinear simulations \cite{DavidIanReview}. The interaction of different particle species through the hot, turbulent and neutrino-rich phase of a neutron star merger means that the underlying equations of motion are far too complex for any current numerical simulation. The LES framework provides a way of incorporating and modelling these features through minor modifications to a single perfect fluid model, even if these modifications are necessarily phenomenological. 

The current state of the art is represented by two related, but at the same time distinct, efforts. The first is focused on  fluid turbulence \cite{radice2017general,radice_calibrated,radice2023abinitio} while the second aims to model the impact of small-scale dynamics on the large-scale magnetic field (expected to be associated with dynamo action in a post-merger remnant) \cite{vigano2019extension,carrasco2020gradient,vigano2020general,aguilera2020turbulent,aguilera2022universality,palenzuela2022turbulent,palenzuela2022large,aguilera2023role,izquierdo2024large}. Both efforts have led to interesting results and convincing evidence that a well thought through LES strategy is a necessary ingredient for realistic simulations. Having said that, the current approaches may be viewed as pragmatic and each has features one may find somewhat``unattractive''. 

\subsection{Filtering the 3+1 equations}

Consider, first of all, a pure fluid simulation carried out using the standard 3+1 foliation approach. All current efforts to implement an LES strategy for this problem consider (following the work of Radice \cite{radice2017general}) the filtering to be done in the spatial hypersurfaces of the spacetime foliation. This has the advantage of being relatively easy to implement as it involves minimal changes to existing evolution schemes. Unfortunately, this strategy breaks the connection to the microphysics and this could be problematic. 

Let us explain. From the foliation point of view, the evolution equations involve a time coordinate originating from the normal to each hypersurface, $N^a$, distinct from the proper time associated with a given fluid element worldline and the fluid four velocity, $u^a$. In effect, by taking the foliation as our starting point, we are dealing with a different space-time split from that used for the local matter physics and the equation of state which ultimately ``lives'' in the fluid frame.  A consistent model requires a careful connection between the two frames. 
 
 As an explicit example, consider the fluid pressure. It is easy to show that 
 filtering affects the inferred isotropic pressure and necessarily involves foliation dependent quantities. In order to  faithfully represent the actual thermodynamics we need to 
 keep track of how the closure relations that need to be introduced (see later) impact on relations associated with the fluid frame. In short, there is no ``equation of state'' that links foliation quantities. This leads to complicated questions for the translation from evolved (conservative) to primitive variables, which need to be understood if we want to minimise undesirable systematics. 

We have already seen that a filtering operation will generally introduce effective non-ideal pieces in the stress-energy tensor. Schematically---ignoring details like the specific filter kernel, the link to  grid resolution, etcetera---we expect to have 
\begin{equation}
    \la T^{ab}_\mathrm{ideal}\ra = \tilde T^{ab} \;,
\end{equation}
with the right-hand side taking the form from \eqref{Tni}, notably including the relativistic analogue of the classic Reynolds stresses, $\tilde \tau^{ab}$. 
As discussed in \cite{celora2021covariant} and demonstrated explicitly in \cite{Celora2024lagrangian}, filtering an ideal fluid on the micro-scale will always lead to a non-ideal fluid at the meso-level.

We also know that if we filter in the spatial directions given by a spacetime foliation, the result will inevitably depend on the choice of gauge (after all, the relevant projection is built from $N^a$ and hence will be gauge dependent). 
For illustration, we may express this as  
\begin{equation}
    \tilde \tau^{ab} = \tilde \tau^{ab} (\dots, \alpha, \beta^i)
\end{equation}
where $\alpha$ and $\beta^i$ are the lapse and shift of the 3+1 foliation. 
As the functional form of the equation of state is  generically non-linear, residuals will also enter the pressure relation. 
Again, schematically, we may write this as (we will develop the idea further later)
\begin{equation}
    \la p (n, \veps) \ra =  p_\mathrm{EOS}( \tn ,  \teps)  +\tilde M \;, \qquad \tilde M = \tilde M (\dots, \alpha, \beta^i)\;, 
\end{equation}
where $p_\mathrm{EOS}$ may be assumed to inherit the functional form from the micro-model (although now acting on meso-scale quantities).
Equivalently, we may introduce a filtered thermodynamic potential, $\tilde p$, such that
\begin{equation}
    \tilde p =  p_\mathrm{EOS} (\tn, \teps)  = \la p \ra - \tilde M \;, 
\end{equation}
implicitly involving residual terms ($\tilde M$) that need to be modelled. 
The key point is that, if we filter in the foliation these residuals will be gauge dependent as we must have
$\tilde p = \tilde p(\dots, \alpha,\beta^i)$.

This is, in turn, problematic for the conservative to primitive inversion required by any existing numerical implementation of the evolution equations.
This step generally requires evaluating $\tilde p$ using a supposedly known equation of state---or the  pressure potential indicated by $p_\mathrm{EOS}$. 
This is tricky as the equation of state relation is assumed to describe the actual microphysics of matter, and hence should not be affected by the gauge choices. 
Having residuals depending on  gauge  is clearly an undesirable feature, but it may be argued  that this is not too much of an issue as long as the evolution scheme is internally consistent. We emphasize, however, that the implications of the primitive inversion step are problematic for such internal consistency arguments.

The natural solution to the problem, allowing us to stay in closer contact with the micro-scale model, is to build  the averaging/filtering in a suitable fluid frame and then ``translate'' the results into the foliation. This is the strategy advocated in \cite{celora2021covariant,Celora2024lagrangian} and the logic we expand on here.

\subsection{Filtering the fields}

Pausing the development to consider problems involving electromagnetism, the standard approach is to filter/average Maxwell's equations for the electric and magnetic fields \cite{brandenburg2005review}. This is, indeed, the strategy adopted in the recent body of work by Palenzuela and colleagues \cite{vigano2019extension,carrasco2020gradient,vigano2020general}. Unfortunately, these models are also problematic. 

The argument is fairly straightforward.
Ignoring for a moment the point that filtering in the foliation breaks the connection to the microphysics,  let us suppose that we want to filter directly in the foliation and focus on the foliation fields. 
We then start by writing down the Faraday tensor decomposition according to a foliation observer $N^a$ (with a slight abuse of notation that should not cause confusion):
\begin{equation}
    E_a = F_{ab}N^b \;, \qquad B_a = \veps_{dabc}N^d F^{bc} \;.
\end{equation}
Assuming that the foliation observer is ``transparent'' to the filtering operation (see later), we then have
\begin{equation}
    \la E_a \ra = \la F_{ab}\ra N^b \;,\qquad  \la B_a \ra = \frac{1}{2}\veps_{dabc}N^d\la F^{bc}\ra \;,
\end{equation}
and we can reconstruct a meso-scale Faraday tensor $\tilde F_{ab}$ from these fields in the usual way
\begin{equation}
    \tilde F_{ab} = N_a \la E_b\ra - N_b \la E_a\ra + \veps_{dabc}N^d \la B^c \ra \;.
\end{equation}
It is now a simple exercise to show that $\tilde F_{ab} = \la F_{ab}\ra$.  The Faraday tensor reconstructed from the filtered foliation fields is the same as the one obtained from filtering the Faraday tensor directly. This result clearly relies on the \emph{assumption} that the metric and the gauge fields vary on larger scales and hence remain unaffected by the filtering operation, but suggests that if we choose to ignore the intrinsic issues with filtering in the 3+1 foliation directly, the Maxwell sector appears not to introduce additional problems.

The story changes when we turn to approximations. In magnetohydrodynamics, for example, we assume that the electric field rapidly adjusts to ensure local (quasi-)charge neutrality \cite{andersson2022physics}. 
The typical argument involves decomposing the filtered charge current according to some observer $U^a$ 
\begin{equation}
    \la j^a \ra = \sigma U^a + J^a \;,
\end{equation}
and then stating that the charge density, $\sigma$, vanishes when $U^a$  corresponds to a locally co-moving observer ($u^a$, say).
At this point we run into trouble. If we filter in the foliation, every filtered quantity will (again) be gauge dependent. In particular,
\begin{equation}
    \la j^b \ra = \la j^b \ra (\dots, \alpha, \beta^i) \Longrightarrow \sigma = \sigma(\dots, \alpha, \beta^i) \;.
\end{equation}
As before, we face problems  when we try to make a connection with the microphysics. There is no easy way to argue that $\sigma$, which depends on the gauge choices, is related to a faithful micro-scale quantity---the charge density as measured by a local observer---and therefore  vanishes. The argument for the MHD approximation becomes artificial.

The  natural way to deal with this problem is to accept the simple fact that the electric and magnetic fields are observer dependent quantities.  This should be enough for us to appreciate that applying the filtering operation, whatever it may be, directly on the electric and magnetic fields will  lead to trouble. 
For example, we know from the discussion in \cite{celora2021covariant} that $\la u^a\ra$ is not going to be normalized, meaning that the filtered four velocity cannot serve as an adequate observer. 
Moreover, even if we ignore this point, we note that 
\begin{equation}
    \la E_a \ra = \la F_{ab}u^b\ra \neq \la F_{ab}\ra \la u^b\ra \;,
\end{equation}
and similarly for the magnetic field.
In essence, because the definitions of the electric and magnetic fields are nonlinear (involve products of tensors), their meaning before and after filtering is not the same. 

Focusing on this point and working with
 a new observer $\tilde u^a$ at the filtering level\footnote{Later we will take this to be a Favre-weighted observer such that $\la n^a\ra = \tilde n \tilde u^a$, but at this point this does not matter.}, we   
introduce the filtered electric and magnetic fields---with respect to this observer---as obtained from the filtered Faraday tensor, and introduce explicitly the decomposition from \cref{eq:FaradayDecomp}. 
This exercise leads to\footnote{As a consistency check, note that if we ignore the filtering operation and consider $\tilde u^a$ to be the same as $u^a$, then the conditions $u^au_a = -1,\,u_a E^a = u_a B^a= 0$ give us $\tilde E^a = E^a$ and $\tilde B^a = B^a$.}
\begin{subequations}
\begin{align}
    \tilde E^a &= \tilde u^b \la u_a E_b \ra - \tilde u^b \la E_a u_b\ra - \frac{1}{2}\varepsilon_{dabc}\tilde u^d \la u^b B^c\ra \;, \\
    \tilde B^a &= \tilde u^b \la u_a B_b\ra -  \tilde u^b \la B_a u_b\ra  + \frac{1}{2}\varepsilon_{dabc}\tilde u^d \la u^b E^c\ra \;.
\end{align}
\end{subequations}
These expressions are all we need to pinpoint  the problem. 
If we assume that $u^a$ is an observer locally co-moving with the fluid, and that the charge-neutrality assumption holds (locally), then we would expect $E_a = 0$ in the micro-model. 
This does not, however, translate into the electric field vanishing at the filtered level, on the meso-scale. Instead, we would have
\begin{subequations}
\begin{align}
    \tilde E_a &= - \frac{1}{2}\varepsilon_{dabc}\tilde u^d \la u^b B^c\ra \; , \label{Etfield}\\ 
    \tilde B_a &= \tilde u^b \la u_a B_b\ra -  \tilde u^b \la B_a u_b\ra  \;.
\end{align}
\end{subequations}
If we insist on working with the filtered magnetic field $\la B^a\ra$, each of these terms requires the introduction of some kind of closure relation.
As a simple matter of fact (demonstrated later), any LES filtering strategy necessarily turns ideal MHD into non-ideal electromagnetism.

A natural way to avoid the issues that arise as we  filter observer dependent quantities is to work with the Maxwell equations at the ``higher level'' of the Faraday tensor (or, equivalently, the vector potential). In the same spirit, the Lagrangian filtering in \cite{Celora2024lagrangian} is performed directly on the stress-energy tensor. This will not resolve the issue of the loss of charge neutrality or breaking the MHD assumptions, but it will remove the need for some of the closure relations indicated above. This is attractive as it simplifies the modelling.

\section{A higher-level strategy}

Before we proceed, and to keep the discussion in the proper context, it is important to acknowledge an unavoidable fact: All averaged/filtered fluid models are wrong! With this in mind, one could opt to ignore the objections we have raised, keep calm and carry on. However, it seems reasonable to ask if we can somehow do ``better''. At the very least, it would be sensible to try to understand what the issues are and how they impact on the physics we are trying to model. This is, indeed, the spirit that motivates the ``higher-level'' (covariant+Lagrangian) filtering strategy outlined here (expanding the discussion in \cite{celora2021covariant} and connecting with the practical demonstration in \cite{Celora2024lagrangian}). 

First of all, recall that discussions of current implementations of LES schemes for general relativistic fluid simulations regularly refer to the fact that the chosen closure relations are gauge dependent. The veracity of this should be evident from the arguments we have provided.  However, the statement is at odds with the model outlined in \cite{celora2021covariant,Celora2024lagrangian}, where the closure relations were discussed via the introduction of an observer (in turn linked to ``average'' properties of the flow). With this Lagrangian strategy it is possible to construct an LES model that is independent of the foliation gauge required by a numerical simulation. This ``gauge invariance'' is key to ensuring that the model is covariant, a desirable property for any truly relativistic scheme. 

\change{On the practical side, covariance also allows us to disentangle as much as possible the ``explicit'' filtering aspects from the ``implicit'' ones: this is so because the filtering operation here does not depend on the coordinates chosen, and therefore on the discretization required by any numerical implementation. 
This does not mean, however, that a model calibrated this way is completely independent from the specifics of the numerical scheme, and as such `a-posteriori' tests are crucial to validate the model performance.}

In addition to promoting the role of the filtering observer, we need to make sure that the filtering operation does not affect the spacetime metric. Otherwise we open a can of worms we would sensibly leave untouched. For example,  we would have
\begin{equation}
    \la E_a \ra = \la g_{ab} E^b \ra \neq \la g_{ab} \ra \la E^b \ra \ ,
\end{equation}
which would be problematic. A similar argument applies to the filtered covariant derivative and the Christoffel symbols. While it is entirely possible that one might be able to come up with an interpretation such that the filtered metric quantities are associated with some ``alternative theory of gravity'', this is not a very attractive proposition. If the actual aim is to model fluid turbulence, extending gravity beyond Einstein may be biting off more than we are prepared to chew. As discussed in \cite{celora2021covariant}, the natural way to formulate rigorously the idea that the metric varies on larger scales naturally leads us to the use of Fermi coordinates \cite{GravitationMTW}.
With the Fermi-coordinate construction the filtered Einstein field equations become
 \begin{equation}
     \la G_{ab} \ra = G_{ab} = 8\pi \la T_{ab}\ra \;,
 \end{equation}
 and the fluid equations of motion can be taken to be
 \begin{equation}
     \nabla_a \la T^{ab}\ra = 0 \;.
     \label{stcon}
 \end{equation}
The fact that the geometric sector of the theory is unaffected by the filtering operation  also has a practical consequence: it allows us to apply the scheme to special relativistic numerical data and lift the results of our analysis to any spacetime. 
 
Another key aspect of a covariant approach to the filtering problem is the need to filter tensorial objects like, $\la T^{ab}\ra$ or $\la F^{ab}\ra$, rather than their projections. 
We have already seen an example of this, as the interpretation issues that arise with filtering the electric and magnetic fields are intimately related to this more mathematical aspect. 
Similarly, in the practical demonstration in \cite{Celora2024lagrangian} we filtered the stress-energy tensor rather than the individual spatial fluid velocities.

\subsection{Filtered fluid dynamics} 

Promoting the view that it is preferable to filter in a suitable ``fluid frame'' in order to connect with the micro-model, it is natural to build the construction from the equation for baryon number conservation. We immediately have
\begin{equation}
\la \nabla_a n^a\ra = \nabla_a \langle n^a \rangle = 0  \ ,
\label{bary3}
\end{equation}
but hit a snag as soon as we want to identify the baryon number density. In general, it will be the case that
\begin{equation}
\la n^a \ra = \la n u^a \ra \neq \la n \ra \la u^a \ra \ , 
\end{equation}
and it is also the case that the filtered four velocity $\la u^a\ra$ does not retain the normalisation one would require from a suitable ``observer'' in relativity \cite{celora2021covariant}. We now have a choice. We can ignore the normalisation issue and introduce a closure relation such that 
\begin{equation}
    \la n u^a \ra = \la n \ra \la u^a \ra + \mathcal N^a \ ,
\end{equation}
with some suitable form for
\begin{equation}
    \quad \mathcal N^a = \la n u^a \ra - \la n \ra \la u^a \ra \ .
\end{equation}
However, we do not find this option particularly attractive. None of the interpretations of $\mathcal N^a$, being either associated with  non-conservation of $\la n\ra$ (as indicated in  \eqref{tdna}) or representing a drift of the filtered baryon density with respect to the chosen observer, are appealing. A more natural strategy would be to introduce a suitably weighted observer $\tilde u^a$---in the spirit of the Favre-weighting commonly used in non-relativistic problems \cite{SchmidtLES}---such that 
 $\tilde u_a \tilde u^a=-1$ and 
 \begin{equation}
    \tilde n = - \tilde u_a \langle n^a \rangle \ .
\end{equation}
We then have 
\begin{equation}
    \nabla_a \la n^a \ra = \nabla_a (\tilde n \tilde u^a) = 0 \ .
\end{equation}
Moreover, we may introduce the projection
\begin{equation}
\tilde \perp^a_b = \delta^a_b + \tilde u^a \tilde u_b \ ,
\end{equation}
allowing us to work out the energy and momentum equations from \eqref{stcon} in the usual way. These equations will, however, require the introduction of  closure relations. 

Starting from a perfect fluid micro-model we can write the filtered version of the stress-energy tensor  as (noting the element of choice associated with the closure term $\tilde{\tau}^{ab}$)
\begin{equation}
\langle T_\text{F}^{ab} \rangle 
= (\langle p \rangle + \langle \varepsilon \rangle ) \tilde u^a \tilde u^b + \langle  p \rangle g^{ab} + \tilde\tau^{ab} \ .
\label{filtab}
\end{equation}
In this case, the closure involves finding a workable representation for 
\begin{equation}
\tilde \tau^{ab} =  \langle ( p + \varepsilon)  u^a  u^b \rangle - ( \langle p \rangle +  \langle \varepsilon \rangle ) \tilde u^a \tilde u^b  \ .
\end{equation}

Evidently, the energy density measured by $\tilde u ^a$ is given by
\begin{equation}
    \tilde \varepsilon = \tilde u_a \tilde u_b \langle T_\mathrm{F}^{ab} \rangle = \langle \varepsilon\rangle + \tilde u_a \tilde u_b \tilde \tau^{ab} \ , 
\end{equation}
Explicitly, we then have
\begin{equation}
\label{eq:filtered_tab}
\langle T_\mathrm{F}^{ab} \rangle 
= (\langle p \rangle + \tilde \varepsilon  ) \tilde u^a \tilde u^b + \langle  p \rangle g^{ab} + 2 \tilde u^{(a} \tilde q^{b)} + \tilde s^{ab} \ , 
\end{equation}
with
\begin{equation}\label{eq:filtered_tab_q}
    \tilde q^a = - \tilde \perp^a_b \tilde u_c \tilde \tau^{cb} = - \tilde \perp^a_b \tilde u_c \,\langle ( p + \varepsilon)  u^c  u^b \rangle\ , 
\end{equation}
and
\begin{equation}\label{eq:filtered_tab_sab}
    \tilde s^{ab} = \tilde \perp^a_c \tilde \perp^b_d \tilde{\tau}^{cd} = \tilde \perp^a_c \tilde \perp^b_d \, \langle ( p + \varepsilon)  u^c  u^d \rangle \ .
\end{equation}

These expressions bring out the expectation that the filtered/averaged stress-energy tensor, when written with respect to the chosen filtered/averaged observer four-velocity, can be interpreted as a non-ideal fluid---not necessarily in thermodynamical equilibrium and for which the second law of thermodynamics need not apply. This is generally the case for LES models. To some extent it leads us into unfamiliar territory, but---depending on how we decide to close the LES scheme---we may draw on the vast body of work related to the stability of relativistic models for dissipative fluids, using the techniques of Hiscock and Lindblom,  Bemfica-Disconzi-Noronha-Kovtun, and  Israel and Stewart \cite{ISRAEL1976,transientStewart77,IsraelStewart79,IsraelStewart79bis,Bemfica19,Bemfica19MIScausal,Bemfica2018,Bemfica2020,KovtunStable,HoultKovtun2020,rocha2024theories,Salazar2019,LivRev_NilsGreg,FrontiersGavassinoAntonelli,Gavassino:2020ubn,Gavassino:2021kjm,Wagner:2023jgq,BVinSim}.

From a numerical implementation point of view, it is appealing to treat an LES model as the minimal changes necessary to the simulation code (and hence to the formalism and how it is interpreted) in order to represent (the main features of) sub-grid turbulence. In effect, we want to interpret the equations as being those of an ideal fluid in thermodynamic equilibrium---viewed in the Eckart frame associated with (in our case) $\tilde u^a$---with correction terms. These correction terms are assumed to not change the meaning, or interpretation, of any thermodynamic quantities.
Unfortunately, these two viewpoints on LES are not quite compatible. It is well known that Eckart-frame non-equilibrium hydrodynamics is unstable \cite{HiscockLindblom1983,HiscockInsta}. Yet, LES schemes have been successfully used in relativity. So, what is going on, and can we really trust these schemes? In order to understand the answer we need to dig a little bit deeper. 

In essence, our previous work  \cite{celora2021covariant} illustrates how LES models can be interpreted as effective field theories. In particular, when the micro-scale theory is itself a single perfect fluid---as in our example here---effective viscosities and heat flux closures must be provided, together with an interpretation of the meaning of the filtered meso-scale fluid four-velocity. The implementation of an LES model then becomes technically identical to that of a non-ideal relativistic fluid, although practical implementations often do not treat it this way. 

One may ask whether the interpretation of an LES model as a non-ideal fluid really matters when these models are  implemented as minor modifications of a perfect fluid. In Newtonian theory the wealth of experience suggests there are no problems with this. However,  the well known problems of first-order non-ideal relativistic fluids suggest the situation may be different and we need to be mindful of the issue. 
A similar question links to stability. Linear stability---the minimal requirement for a useful model---has been explored in \cite{duez2020comparison,celora2021covariant}.
From a pragmatic point of view, this issue is related to whether unstable wavenumbers are resolved, or not, in the coarse-grained simulation---an aspect  discussed in the context of bulk-viscous simulations in \cite{BVinSim}.
However, nonlinear stability is more complex. The interpretation as a non-ideal fluid is helpful, as the entropy current can be used to discuss entropy stability of the full system, but much more work is required if we want to understand this problem.

\subsection{Filtered thermodynamics}

In general, LES filtering  also impacts on the thermodynamics and the equation of state. This is already evident from \eqref{Gibbs}. 
At the micro-level,  thermodynamics provide  a direct link between the pressure and the number and energy densities, e.g. $p = p (n,\veps)$, representing the matter equation of state. This relation is ``confused'' by  filtering. The filtered stress-energy tensor in \cref{eq:filtered_tab} involves $\la p\ra$, but the fact that \eqref{Gibbs} involves non-linearities implies that  $\la p \ra = \la p (n,\veps)\ra\neq p (\tilde n, \teps)$. We could choose to ignore this issue---not consider any additional closure/residual terms---but the results from  \cite{Celora2024lagrangian} demonstrate that this can introduce undesirable systematic errors in the equation of state parameters. This  may  be problematic, so it makes sense to ask if we can do better. At the very least, we need to understand the problem.

It is easy to see that any filtering operation  involves breaking the link between the micro-model equation of state (which tends to be based on nuclear physics arguments) and the meso-model. This is immediately clear from the filtered version of \eqref{Gibbs}, which (obviously) takes the form
\begin{equation}
    \la p \ra = - \la \veps \ra + \la n\mu\ra \;.
\end{equation}
Taking the view that it is natural to work with $\tn,\teps$ as defined/measured by the Favre observer on the meso scale, we may rewrite this as
\begin{equation}
\la p \ra = - \tilde \veps + \tilde n \tilde \mu + \mathcal M \;,
\end{equation}
with
\begin{equation}
    \mathcal M = \la n\mu\ra - \tilde n \tilde \mu + \tilde \veps - \la \veps\ra =  \la n\mu\ra - \tilde n \tilde \mu + \tilde u_a \tilde u_b \tau^{ab} \;.
\end{equation}
At this point we have options. The simplest choice would be to assume that, if the micro-model is built from a single parameter, $\veps = \veps(n)$, then the meso-model should be a single-parameter model, as well. Effectively, this would involve starting from a functional $\tilde \veps= \tilde \veps(\tilde n)$, assuming that the meso-level chemical potential is defined in the usual way and then defining the meso-scale pressure as
\begin{equation}
\tilde p = - \tilde \veps + \tilde n \tilde \mu \;.
\end{equation}
That is, we can let the pressure potential ``soak up'' the closure aspects by defining
\begin{equation}
    \mathcal M = \la p \ra - \tilde p \;. 
\end{equation}
However, it is evident from the construction that we should not necessarily expect the new equation of state $\tilde p (\tilde n)$ to take the same form as in the micro model. We can assume that this is the case---as, indeed, tends to be done in numerical simulations---but we need to keep in mind that  this involves imposing a specific closure model.

An alternative interpretation---perhaps natural given that the filtering of an ideal fluid leads to non-ideal fluid equations---would be to assume that the filtered energy no longer follows from a single parameter. Instead, filtering could be assumed to introduce an effective (not physical) entropy, $\tilde s$. Then, starting from $\tilde \veps(\tilde n, \tilde s)$, we can work out the associated ``temperature''
\begin{equation}
    \tilde T = \left( \frac{\partial \tilde \veps}{\partial \tilde s} \right)_{\tilde n} \;,
\end{equation}
(obviously depending on the chosen energy functional...) and let the pressure follow from
\begin{equation}
\tilde p = - \tilde \veps + \tilde n \tilde \mu + \tilde s\tilde T \;.
\end{equation}
In effect, we would now be taking the view that the closure terms are given by
\begin{equation}
    \mathcal M = \la p \ra - \tilde p  = \la \mu n\ra - \tilde \mu \tilde n - \tilde T \tilde s + \tilde u_a \tu_b \tau^{ab}
\end{equation}
This model is quite different from the single-parameter one, illustrating the fact that the meso-model equation of state involves aspects that are somewhat arbitrary.
Despite this inevitable degree of arbitrariness, however, filtering at the higher-level allows us to work around the discomforting issue of having equation of state residuals depending on the specific choice of gauge.

It is worth keeping in mind that the problem is ultimately statistical in nature. The practical demonstrations in \cite{Celora2024lagrangian} highlight the fact that, in reality, the energy $\tilde \veps$ would be obtained as a distribution in terms of $\tilde n$. This means that the closures/residuals will also be distributions. This, in turn, allows for a broader range of functional forms for $\tilde \veps$ and more freedom in the specification of the equation of state model on the meso-scale. This inevitably means that the problem becomes rather complicated.

The freedom in specifying the thermodynamic quantities for the filtered  meso-model must be accounted for in a numerical implementation. This suggests  a  potential tension. We want to benchmark simulation results against (say) observed gravitational-wave signals and extract information regarding the supranuclear equation of state. 
However, numerical simulations involve an effective filtered model.  In particular, the LES scheme comes to the fore when the flow is turbulent, e.g. in an actual neutron star merger event, while it does not have substantial impact when the flow is laminar, e.g. during the inspiral phase of binary evolution when the two stars  are well separated. Due to the impact of the filtering closure terms, these two phases of the  evolution effectively involve different ``equations of state'' for the same micro-model. The question  of how much this discrepancy contributes to systematic errors and bias in parameter extraction has so far been largely ignored. This may be  problematic.

\subsection{Entropy pairs}

The impact of filtering on the meso-level interpretation of the entropy could be important for numerical stability as well as physical accuracy. An essential way of discussing nonlinear numerical stability in systems that admit shocks (like neutron star mergers) is that of \emph{entropy pairs}: a positive definite scalar function that is a convex function of the state vector, satisfying (weakly) a conservation law, whose existence guarantees the existence of weak solutions. Even when the full system is not in balance law form, the existence of an entropy pair allows for a rigorous construction of shock solutions, and its properties determine the shock speeds of the system.

Despite the name, entropy pairs need have nothing to do with physical entropy. However, in fluid problems they are often tightly linked. For fully general relativistic  hydrodynamics no entropy pair has been computed, but work in special relativity (see e.g.~\cite{Wu2020}) suggests that the specific entropy current $s^a$ will act (after projection into the numerical coordinate system) as an entropy pair, as expected\footnote{Note that entropy pairs are not unique: in many systems there are an infinite number of choices of entropy pairs. For balance laws this is not important: all allowable pairs give the same stability results, and the shock solutions are the same. When the system cannot be written in balance law form then the form of the entropy pair preserved by the numerical scheme determines the shock speed constructed, and the choice of the ``physical'' entropy pair has direct, physical impact.}.

When interpreted in the context of filtering, this highlights the importance of the \emph{choice} of the meso-entropy current $\tilde{s}^a$, linked to the choice of residual closure terms $\tilde{s}, \tilde{T}$. In order to act as an entropy flux $\tilde s^a$ has to satisfy three crucial criteria. First, the projection $N_a \tilde{s}^a = \tilde{S} \sim \tilde{s}$ must be positive. Second, the projection $\tilde{S}$ must be convex as a function of the evolved variables of the system (typically $N_a n^a$ and\footnote{We here indicate by $e^{(j)}_b$ the j-th spatial leg of the foliation tetrad.} $N_a e^{(j)}_b T^{ab}$). Finally, at the continuum level the balance law 
\begin{equation}
    \nabla_a \tilde s^a \le 0
\end{equation}
should be satisfied, meaning that entropy only changes at discontinuities, and always decreases\footnote{The sign is a conventional choice linked to the fact that, in the strong form, this function is usually considered as an energy. Strictly, the standard entropy pair uses $-s^a$ or equivalent to flip the sign.}. The first two criteria are directly linked to the choice of the residuals in filtering, and can be interpreted in terms of the thermodynamic stability of the meso equation of state in terms of the meso thermodynamic potentials.

\subsection{Filtered electromagnetism}

Turning again to problems involving electromagnetism, the advantage of filtering at the higher level immediately becomes apparent. Expressed in terms of the Faraday tensor $F_{ab}$ (or, should we prefer, the associated vector potential $A_a$) the covariant form of Maxwell's equations is linear. From \eqref{eq:MaxwellCovariant} we immediately have\footnote{The same is true if we opt to work with the vector potential: given the properties of the Lagrangian filtering discussed in \cite{celora2021covariant}
$$
\la F_{ab} \ra = \la \nabla_a A_b - \nabla_b A_a \ra = \la \nabla_a A_b\ra  - \la \nabla_b A_a \ra =  \nabla_a \la A_b\ra  -  \nabla_b \la A_a \ra \ .
$$
} 
\begin{equation}
    \nabla_a \la F^{ba} \ra = \mu_0 \la j^b\ra \;,
    \end{equation}
    and
    \begin{equation}
     \nabla_{[a} \la F_{bc]}\ra  = 0 \;.
\end{equation}
Moreover, given the Favre-weighted observer we can 
define new fields such that
\be 
\langle F^{ab} \rangle = \tilde u^a \tilde E^b - \tilde u^b \tilde E^a + \epsilon^{abcd} \tilde u_c \tilde B_d \;,
\ee
with 
\be 
\tilde E^a = \tilde u_b \langle F^{ab} \rangle \;,
\label{Edef}
\ee
and
\be 
\tilde B_a = -\frac{1}{2} \epsilon_{abcd} \tilde u^c \langle F^{cd} \rangle \;.
\label{Bdef}
\ee
Formally, the filtered version of Maxwell's equation now retains the usual form, even if we decide to work with the filtered fields $\tilde E^a$ and $\tilde B^a$. This is attractive. 

Moreover,  it is  easy to see that a covariant filtering operation---that commutes with the covariant derivatives---will preserve the imposed gauge conditions as long as these are linear. 
This would, for example, be  the case for the familiar Lorenz gauge \cite{Etienne2012_EM_gauge}:
\begin{equation}
    \nabla_a A^a = 0 \Longrightarrow \nabla_a \la A^a \ra = 0 \;.
\end{equation}
It is, however, not the case for the ``generalized'' version that tends to be used in numerical simulations (see, for example \cite{Cipolletta2019Spritz}). In effect, we would have \cite{PhysRevLett.109.221102}
\begin{equation}
    \nabla_a A^a = \xi N^a A_a \;,
\end{equation}
with a notably nonlinear and gauge-dependent right-hand side ($\xi$ is a fixed parameter). 
In this case, we clearly have to be more careful, as the gauge condition is preserved only as long as $N^a$ is transparent to the filtering operation. 
At this point it seems natural to argue that this ought to be the case as long as the normal to the foliation varies on the metric scales.

There is, of course, no such thing as a free lunch. Filtered electromagnetism still requires phenomenological closure relations. First of all, let us consider the charge current. We can always define
\be
\tilde j^a = \langle j^a \rangle = \tilde \sigma \tilde u^a + \tilde J^a  \ , 
\ee
with
\be
\tilde \sigma = - \tilde u_a \langle j^a \rangle \;,
\ee
and
\be 
\tilde J^a = \tilde \perp^a_b \langle j^b\rangle  \;.
\ee
We should also have
\be 
\langle j^a \rangle = \langle \sigma u^a \rangle + \langle J^a \rangle \;,
\ee
from which it follows that
\be 
\tilde \sigma = - \tilde u_a \left( \langle \sigma u^a \rangle + \langle J^a \rangle\right) \;.
\ee
From this argument it is clear that, even if the micro-scale flow is locally charge neutral (in the sense that $\sigma = 0$) this may not be true for the filtered flow. Instead, we get
\be
\tilde \sigma = - \tilde u_a \langle J^a \rangle \;,
\ee
which does not have to vanish. Conversely, if we impose charge neutrality on the filtered flow then this condition may not hold on the micro-scale. This would be problematic given that nuclear physics arguments tend to  impose charge neutrality at the level of the equation of state. Accepting that the filtered system may not be charge neutral \cite{andersson2022physics} we (again) need to pay attention when we carry out the equation of state inversion in a numerical simulation. 

A related issue arises in ideal MHD, which typically involves an assumption of high enough conductivity that the electric field can be taken to vanish in the fluid frame. For the present model, it is evident from \eqref{Etfield} that the filtered electric field will typically not vanish. The equations we arrive at by filtering MHD will be the general equations of electromagnetism. In fact, it may not make much sense to first make the MHD-approximation and then apply the filtering procedure.  The main reason for this has to do with the choice of filtering observer and the simple fact that the electric and magnetic fields are observer dependent. That such a strategy would be problematic is evident given that the issue of local charge neutrality and vanishing electric field are linked.
One may, of course, impose the ideal MHD conditions on the filtered system but this is also problematic as it implies that the same conditions do not hold on the micro-scale (we will make this explicit later). 

So far, we have not had to introduce closure relations for the filtered Maxwell equations. This changes when we consider, first, the stress-energy tensor and the Lorentz force and, second, Ohm's law and resistivity. 

In order to specify the meso-model we need to make choices. The model will generally involve some combination of non-ideal hydrodynamic/electromagnetic terms. The balance between these terms is likely to be different in different large-eddy implementations. For example, focussing first of all on the 
Lorentz force, we immediately get (assuming that it makes sense to split the filtered stress-energy tensor in the same way as in the micro-model)
\be 
\nabla_a \langle T^{ab}_\mathrm{F} \rangle = - \nabla_a \langle T^{ab}_\mathrm{EM} \rangle = \langle f^b_\mathrm{L} \rangle \;,
\ee
with
\be
\langle f^b_\mathrm{L} \rangle= -
 \tilde j_a \langle F^{ab} \rangle+
\mathcal F^b \ ,
\ee
where
\be
\mathcal F^b =  - \langle j_a F^{ab} \rangle + \tilde j_a \langle F^{ab} \rangle \;,
\ee
has to be represented by a closure relation.

In ideal magnetohydrodynamics, the charge current is obtained from the Maxwell equations. Intuitively this is somewhat back-to-front. The charge current should source the electromagnetic field, rather than the other way around. Moreover, in order to deal with non-ideal aspects, like resistivity, the standard approach is to introduce some version of Ohm's law to provide the electric field. Again, this is a---potentially misleading---simplification. A proper derivation would start the discussion of resistivity at the level of a charged multi-fluid model---noting that the charge current represents an additional fluid degree of freedom, the flow (or perhaps rather, drift) of electrons relative to protons---and derive the MHD equations as the single-fluid limit of the system \cite{NilsPRD12,andersson2022physics}. 
The underlying multi-fluid nature is also important for the development of consistent models beyond ideal. 
This fact motivates work  connected to the variational (dissipative) multi-fluid framework \cite{NilsPRD12,AnderssonCQG2017VariationalPlasmas,BeyondIdealMHD} as well as linking to ``3+1 formulations'' for numerical implementations (see, in particular, \cite{BeyondMHD3+1,andersson2022physics}). From a theory perspective, it is clear that a 
filtering strategy for electromagnetism ought to account for the multi-fluid aspects. In practice, however, this may be too ambitious (given our current understanding of the problem and the simple fact that there have so far been no attempts to filter relativistic multi-fluid systems).  

In absence of a filtered multi-fluid model, we are forced to consider Ohm's law as a phenomenological relation---linking the charge current to the electric field---allowing us to close the system. The classic example of this is the work by 
Bekenstein and Oron \cite{BekensteinOron} which posits that
\be
\perp^a_b j^b = \eta F^{ab} u_b = \eta E^a
\label{ohm}
\ee
where $\eta$ is the conductivity (and $1/\eta$ gives the resistivity).
Since MHD models involve a low-frequency approximation \cite{andersson2022physics}, and considering LES in the sense as a low-pass filtering operation \cite{Berselli2006mathematicsLES},
it would be natural for the filtered model to start from 
\be
\tilde u_b \langle F^{ab} \rangle - \frac{1}{\eta}\tilde \perp^a_b \tilde j^b = \mathcal W^a
\label{ohmfil}
\ee
or
\be
\tilde E^a = \mathcal R \tilde J^a + \mathcal W^a
\ee
with $\mathcal R = 1/\eta$.
We may then introduce an algebraic closure expressed in a basis involving (say) the charge current and the magnetic field. 

With 
\begin{equation}
    \mathcal W^a = \alpha \tilde B^a + \beta \tilde J^a + \gamma \epsilon^{abcd} \tilde J_b \tilde u_c \tilde B_d 
\end{equation}
we would end up with 
\be
\tilde E^a = \tilde{\mathcal R} \tilde J^a + \alpha \tilde B^a + \gamma \epsilon^{abcd} \tilde J_b \tilde u_c \tilde B_d 
\ee
where $\tilde{\mathcal R} = \mathcal R +\beta$. In this expression, the first term represents an effective resistivity, the second term can be taken to represent the classic alpha-dynamo (in the spirit of \cite{BucciantiniDelZanna2013}) while the third term provides an effective Hall term. 
This kind of closure, sometimes referred to as a ``fully covariant mean-field dynamo closure'' has been recently investigated in \cite{Most23_dynamo}. 
Notably, with the approach we advocate for here, we would be able to calibrate the relevant parameters of the model (say $\alpha$) in a fully-covariant fashion using special relativistic simulations of turbulence.

\subsection{Modelling/interpretation issues}

At this point it should be clear that any LES filtering strategy involves a set of choices, potentially impacting on the interpretation of the model. As an illustration of this, let us consider the coupled fluid-Maxwell problem at the meso-level. First of all, we have the energy equation, which (making use of the previous results) takes the form
\be
\tilde u^a \nabla_a \tilde\veps + (\tilde p + \tilde \veps) \nabla_a \tilde u^a =  \tilde J_a \tilde E^a + \tilde u_b \left( \nabla_a \tilde\tau^{ab}  -  \mathcal F^b  \;\right) . 
\label{eneq}
\ee
Evidently, the conversion of energy from fluid to magnetic depends on the closure conditions (e.g. the $\tilde u_b \mathcal F^b$ term).
Meanwhile, the momentum equation can be written
\be
(\tilde p  + \tilde \varepsilon  ) \tilde u^a  \nabla_a \tilde u^c + \tilde \perp^{ac} \nabla_a \tilde p  =  \tilde \sigma \tilde E^c + \left(\epsilon^{ecad} \tilde u_e\right) \tilde J_a \tilde B_d  - \tilde \perp^c_b \left( 
 \nabla_a \tilde \tau^{ab} - \mathcal F^b\right) \;.
\ee

These interpretation issues also have direct impact on practical modelling. 
For instance, if we insist on the residuals having a non-ideal fluid interpretation, then we would look for a model that involves, say, the fluid velocity gradients. 
If we take the opposite view, then we would aim to build a model in terms of, say, the electromagnetic field and the charge current. 
In this sense, numerical experiments and data-driven tests will be useful to assess the suitability of the different options we have listed above, in addition to possible physics constraints. 
More work needs to be done in this direction, but the practical demonstration in \cite{Celora2024lagrangian}, at least, suggests that this is a viable strategy. 

A useful illustration follows by asking if we may introduce closures that makes the equations look like the ones from ideal MHD. In the first instance, we  would then need to remove the electric field from the fluid equations. We can easily do this by making sure that 
\be
\tilde u_b \mathcal F^b = \tilde J_a \tilde E^a \;,
\ee
and
\be
\tilde \perp^c_b \mathcal F^b  =- \tilde \sigma \tilde E^c \;.
\ee
In essence, we need
\be
\mathcal F^b = - \left( \tilde J_a \tilde E^a\right) \tilde u^b - \tilde \sigma \tilde E^b \;.
\label{Fclosure}
\ee
It is not clear why one would want to do this, but it is clearly possible. Having said that, if we keep in mind the implicit aspect of filtering associated with the numerical discretisation and the fact that many simulation codes are based on ideal MHD, then we may be learning something interesting here. 
Any such meso-model is effectively imposing the closure from \eqref{Fclosure}. Implicitly, the micro-model must then be at the level of non-ideal electromagnetism. Whether this is important or not remains to be figured out.

\subsection{Asymptotic preservation}

From a practical point of view, 
current modelling---particularly in numerical simulations---adds a subgrid model to capture physics at scales that cannot be otherwise included. This implicitly---or explicitly---means that an effective theory is solved. The two essential questions are whether the systematic uncertainties in predicted observables from the process are small enough to make the approach useful, and whether the continuum limit of the effective theory is the original theory we want to model.

Our discussion has focused on the first question, by exploring how differing effective theories (meso-models) can be related to the same original theory (the micro-model), and how the impact on potential observables (e.g. equation of state parameters) can be, in principle, quantified.

The second question is linked to the concept of \emph{asymptotic preservation} \cite{Jin_2022}, and can be phrased in this fashion. Suppose model $A$ is the limit of another model, $A_\delta$, as a parameter $\delta \to 0$. Suppose we have a numerical scheme with representative grid-step $\Delta x$. The numerical scheme is asymptotic preserving only if the continuum limit of the scheme tends to the correct solution for the limiting model $A$, \emph{independently} of the order in which $\delta, \Delta x \to 0$.

This understanding is crucial for studying the subgrid models that appear in LES schemes to establish whether the continuum limit is correct. \change{A minimal requirement is that for small enough filtering scales the original system is recovered.} It is \emph{not} sufficient \change{, however,} to note that LES closure models are proportional to $\Delta x$ in order to say that the correct result will be obtained with enough numerical resolution. This only holds true when the scheme is asymptotic preserving. With stiff source terms (which appear in most non-ideal models of MHD~\cite{wright_resistive_2020,PalenzuelaBeyond2009,dionysopoulou2013general,Most:2021rhr} and relativistic hydrodynamics~\cite{Hatton2024DEIFY}, and also in radiation hydrodynamics~\cite{radice2022newM1,2024MNRAS.528.5952M,foucart2023neutrino}) the development of asymptotic preserving schemes is not guaranteed, and the impact of LES schemes on these models is not obvious. Again, there is work to be done. 

\section{Concluding remarks}

A Large Eddy formulation naturally leads to a more general stress-energy tensor than that of ideal matter \cite{celora2021covariant}. As we have discussed, all non-linearities in the micro-model imply the need to introduce (more or less phenomenological) closure relations to complete the modelling at the meso-scale at which numerical simulations are carried out. Given that the purpose of the Large Eddy approach is to capture the maximal physics at minimal additional computational cost, a \emph{practical} numerical implementation will require additional assumptions.

The current state of the art in general relativistic large-eddy modelling is represented by two distinct efforts. 
The first is focused on capturing/modelling the angular momentum transport due to magneto-hydrodynamic turbulence in merger remnants \cite{radice_calibrated,radice2023abinitio}, implementing a Smagorinsky-type closure model \cite{Smagorinksy} calibrated on the high-resolution simulations of \citet{Kiuchi2018super,kiuchi2023SecondLongBNS}. 
The second is aimed at modelling the magnetic field (small-scale) dynamics and amplification, employing a
gradient model with no tunable parameters \cite{vigano2019extension,carrasco2020gradient,vigano2020general,aguilera2020turbulent,aguilera2022universality,palenzuela2022turbulent,palenzuela2022large,aguilera2023role,izquierdo2024large}. As discussed in detail in \cite{DavidIanReview}, both efforts have provided promising results.

However, the large-eddy strategy to the relativistic flows in, for example, merger remnants is still very much at the development stage---particularly when compared to the level of sophistication reached in non-relativistic models---with a number of open issues to be resolved. 
These range from fundamental issues concerning covariance \cite{DavidIanReview,EyinkDrivas2018,celora2021covariant} to interpretation aspects regarding the interplay between filtering and common modelling approximations/assumptions. 
Given that any non-linear closure relation or constraint is affected by filtering, this has implications for the interpretation of the matter equation of state (and the role it plays in a numerical simulation) or, indeed, the notion of electric and magnetic fields.  
The consequences of these issues are profound, as filtering breaks the mathematical link between physical quantities  expected from the micro-physical level. 
In turn, this has repercussions for, for example, the conservative to primitive step key to a numerical simulation, the assumption of charge neutrality central to ideal MHD models. 

We have provided a coherent, although at this point still somewhat  abstract, discussion of the vexing issues that arise in a filtering model.
The scope of the analysis was two-fold: On the one hand, we stressed the implications of any LES scheme, emphasizing  how modelling choices relate to systematics. 
On the other hand, we expanded on \cite{celora2021covariant} and provided  appropriate context for the Lagrangian filtering framework put forward in \cite{Celora2024lagrangian}. 
This framework allows us to explore many of the relevant issues and is specifically designed to be free from  covariance problems. This will, at the very least, allow us to take the next steps towards a better understanding of the problem.

Admittedly, our discussion does not provide precise answers to many of the questions we have raised. However, the intention was to bring to the fore the many varied aspects of the filtering problem and provide a platform for further discussion. Quantitative answers will require specific formulations and (some level of) numerical experimentation (extending the proof-of-principle demonstrations from \cite{Celora2024lagrangian}). Ultimately, the solution to the problem has to be data driven. At the same time, it is important to pay attention to the broader context. For example, we have highlighted the element of choice involved in the modelling. We have also commented on the statistical nature of the problem. These issues  are closely related to the well-known loss of predictability in turbulence simulations~\cite{Lesieur_turbulence}. If the meso-model involves some degree of randomness, then this will be inherited by the output and the inferred observables. Understanding this aspect is important in order to better model the systematic errors in simulations. Moreover, the inevitable link to the problem of uncertainty quantification~\cite{UQchapter} clearly warrants further consideration.
The relevance of this is particularly evident in light of the aim to constrain the supranuclear equation of state using Bayesian inference of observed neutron star dynamics \cite{nedora2021mapping,10.1093/mnras/stab1287}. This aspect of the problem has so far received very little (if any) attention. We  need to do better and  make progress on all aspects of the problem.

\begin{acknowledgments}
TC is an ICE Fellow supported through the Spanish program Unidad 
de Excelencia Maria de Maeztu CEX2020-001058-M.
NA and IH gratefully acknowledge support from Science and Technology Facility Council (STFC) via grant numbers ST/R00045X/1 and ST/V000551/1.

\end{acknowledgments}

\appendix


\bibliography{biblio.bib}

\end{document}